\documentclass[prd,
,superscriptaddress,
nofootinbib,%
tightenlines ]{revtex4}
\usepackage{epsfig}
\usepackage[colorlinks=true,linkcolor=blue,urlcolor=blue,citecolor=blue]{hyperref}
\usepackage{amsfonts}
\usepackage{amssymb}
\usepackage{amsmath}
\usepackage{slashed}
\usepackage{color}

\newcommand{\ben}{\begin{displaymath}}
\newcommand{\een}{\end{displaymath}}
\newcommand{\be}{\begin{equation}}
\newcommand{\ee}{\end{equation}}
\newcommand{\bea}{\begin{eqnarray}}
\newcommand{\eea}{\end{eqnarray}}

\begin{document}
\title{Gravitational form factors of the Higgs boson}
\author{P.~Bei\ss ner}
 \affiliation{Institut f\"ur Theoretische Physik II, Ruhr-Universit\"at Bochum,  D-44780 Bochum,
 Germany}
 \author{B.-D.~Sun}
 \affiliation{Institut f\"ur Theoretische Physik II, Ruhr-Universit\"at Bochum,  D-44780 Bochum,
 Germany}
\author{E.~Epelbaum}
 \affiliation{Institut f\"ur Theoretische Physik II, Ruhr-Universit\"at Bochum,  D-44780 Bochum,
 Germany}
\author{J.~Gegelia}
 \affiliation{Institut f\"ur Theoretische Physik II, Ruhr-Universit\"at Bochum,  D-44780 Bochum,
 Germany}
 \affiliation{Tbilisi State
University, 0186 Tbilisi, Georgia}

\begin{abstract}

We calculate the one-loop electroweak corrections to the gravitational form factors of the Higgs boson and discuss the interpretation of the obtained results. 

\end{abstract}

\maketitle

\section{Introduction}	

After 100 years of quantum mechanics we still continue to use the language of the classical physics to describe subatomic systems. While we certainly do not think that these are in any sense classical objects we still operate with terms like, e.g.,  charge radius, meaning that in low-energy electromagnetic scattering experiments in first approximation they behave like as if subatomic systems were classical objects with charge distributions and the corresponding mean-square radii.    

Being tempted to adopt (a rather problematic) definition of elementary
particles as those which are represented by own local field operators
in the standard model Lagrangian \cite{Weinberg:1996kw} one may ask if
these particles are point-like. This question cannot be answered
unless one specifies the precise meaning of the term point-like.
Probing subatomic systems via electromagnetic interactions, electrons
behave like as if they had  pointlike charge-distribution if one stays
at leading order in the expansion in powers of the electromagnetic
coupling. Protons, on the other hand, possess non-trivial electromagnetic form factors and therefore are seen as extended objects. 
This picture of nucleons occurs due to the strong interaction, which
is much stronger than the electromagnetic one and needs to be taken
into account also when the electromagnetic interaction is treated only
at tree order, i.e., within the classical approximation. 
Similarly to the electromagnetic case, one can, at least in principle,
also probe subatomic systems with gravitons. While it is not feasible
to measure scattering processes of such systems off external
gravitational sources, theoretical investigation can be carried out in
the framework of the standard model of particle physics. Such
scattering processes in the single-graviton approximation are
described by diagrams in which the system under consideration couples
to the energy-momentum tensor (EMT).   
Matrix elements of the EMT and the corresponding gravitational form
factors (GFF) \cite{Kobzarev:1962wt,Pagels:1966zza} have been
extensively studied for various systems in recent years.
In particular, the GFFs of hadrons have attracted much attention, see,
e.g.,
Refs.~\cite{Polyakov:1999gs,Polyakov:2002yz,Polyakov:2018zvc,Ji:1996ek,Hudson:2017xug,Nature,Lorce:2025oot,Kumano:2017lhr,Kumericki:2019ddg,Shanahan:2018nnv,Shanahan:2018pib,Diehl:2006ya,Avelino:2019esh,Belitsky:2002jp,Lorce:2017wkb,Schweitzer:2019kkd,Goeke:2007fp,Polyakov:2020rzq,Alharazin:2020yjv,Gegelia:2021wnj,Epelbaum:2021ahi,Alharazin:2022wjj,Alharazin:2023zzc,Alharazin:2023uhr}. Calculations
of gravitational structure due to strong interaction effects have been
performed using different methods. The increased interest to this
topic is driven by the fact  that GFFs of hadrons can be (indirectly)  extracted from
experiment. 
Compared to the gravitational force,  electromagnetic and weak
interactions are very strong and, therefore, it makes sense to
consider also the electroweak corrections to matrix elements of the EMT.
Corrections due to the electromagnetic interaction lead to infrared
divergences \cite{Kubis:1999db,Freese:2022jlu} that require taking into account emission of soft photons. 
In this work, we calculate the one-loop electroweak corrections to the
matrix element of the EMT operator for the Higgs boson, which is
electrically neutral, and provide interpretation of the obtained results in terms of spatial densities defined via sharply localized states \cite{Panteleeva:2022uii}.

Our paper is organized as follows: In
section~\ref{effective_Lagrangian} we specify some details of the
electroweak theory,  which are needed for our calculation.
Next,  we calculate  in section~\ref{EMFFs} the matrix element of the EMT operator for the Higgs boson and extract GFFs.  
The results of our work are
summarized in  section~\ref{conclusions}.  

\section{Lagrangian and the energy-momentum tensor}
\label{effective_Lagrangian}

In our calculations we use the Lagrangian and the Feynman rules of the electroweak theory as specified in Ref.~\cite{Aoki:1982ed}.
The EMT operator of a non-Abelian gauge theory with spontaneous symmetry breaking 
can be found in Ref.~\cite{Freedman:1974gs}.
We exploited the results of this work to obtain the EMT operator
needed for our calculations by considering the Lagrangian of the
electroweak theory in the presence of the external gravitational field and using the definition \cite{Birrell:1982ix}
\begin{eqnarray}
T_{\mu\nu} (g,\psi) & = & \frac{2}{\sqrt{-g}}\frac{\delta S_{\rm m} }{\delta g^{\mu\nu}}
\label{EMTMatter}
\end{eqnarray}
for matter fields interacting with the gravitational metric fields.
For the fermion fields interacting with gravitational vielbein fields, we employ the definition \cite{Birrell:1982ix} 
\begin{eqnarray}
T_{\mu\nu}  (g,\psi) & = & \frac{1}{2 e} \left[ \frac{\delta S }{\delta e^{a \mu}} \,e^{a}_\nu + \frac{\delta S }{\delta e^{a \nu}} \,e^{a}_\mu  \right] ,
\label{EMTfermion}
\end{eqnarray}
where $e$ is the determinant of $e^a_\mu$. 
 
%

\medskip

In the calculation of one-loop diagrams below we apply dimensional
regularization (see, e.g., Ref.~\cite{Collins:1984xc}) with $D$
spacetime dimensions and use the program
FeynCalc
\cite{Mertig:1990an,Shtabovenko:2016sxi,Shtabovenko:2023idz}. The
results of our calculations are expressed in terms of scalar integrals
defined in appendix  \ref{AppA}.

\begin{figure}[t]
\begin{center}
\epsfig{file=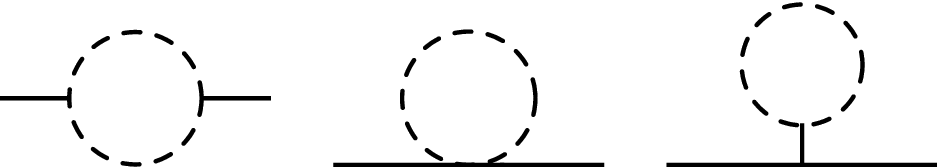,scale=0.5}
\caption{Topologies of one-loop diagrams contributing to the self-energy of the Higgs boson. Solid lines correspond to Higgs bosons and the dashed lines represent vector bosons, fermions, Higgs bosons, Goldstone bosons and Faddeev-Popov ghosts.}
\label{Higgs_SE}
\end{center}
\vspace{-5mm}
\end{figure}

\section{One-loop corrections to the gravitational form factors}
\label{EMFFs}


The one-particle matrix element of the EMT operator for a scalar particle can be parameterized as
\cite{Donoghue:1991qv}:
\begin{equation}
\langle p' | T_{\mu\nu}| p\rangle = \frac{1}{2} \left( g_{\mu\nu} q^2-q_\mu q_\nu \right) \theta_1(q^2) + \frac{1}{2} \, P_\mu P_\nu \, \theta_2(q^2) ,
\label{GFFs}
\end{equation}
where $q=p'-p$ and $P=p+p'$. The gravitational
form factor $\theta_2(q^2)$ satisfies the normalization condition
$\theta_2(0)=1$, while $\theta_1(q^2) $ gives rise to the D-term $D= -
\theta_1(0)$, which is not constrained by normalization.

\medskip

We obtain the Higgs boson matrix element of the EMT operator by applying the standard LSZ formalism to the vacuum expectation value of the time-ordered product of the EMT operator and two Higgs boson fields.  
To do so at one-loop order we first need to calculate the one-loop
contributions to the pole position and the residue of the dressed propagator of the Higgs boson given by
\begin{equation}
S(p) \; =\;  \frac{i}{p^2-M_{H}^2+ \Sigma} \; =\;  \frac{i \, Z}{p^2- z} \; + \; \text{non-pole  part} \,,
\label{HDPr}
\end{equation}
where $M_H$ is the mass of the Higgs boson, $ \Sigma $ is its
self-energy, $z= M_H^2 + \sum_{j=1}^{\infty} \hbar^j \delta z_j $
denotes the pole position of the dressed propagator while
$Z=1+\sum_{j=1}^{\infty} \hbar^j \delta Z_j $ is the residue at the
pole. Here, the  series in terms of $\hbar$ correspond to the loop expansion.
Topologies of one-loop diagrams contributing to the self-energy of the Higgs boson are shown in Fig.~\ref{Higgs_SE}. 
The corresponding expressions for the one-loop contributions to $z$
and $Z$ are given in appendix \ref{AppB}.

\medskip

\begin{figure}[t]
\begin{center}
\epsfig{file=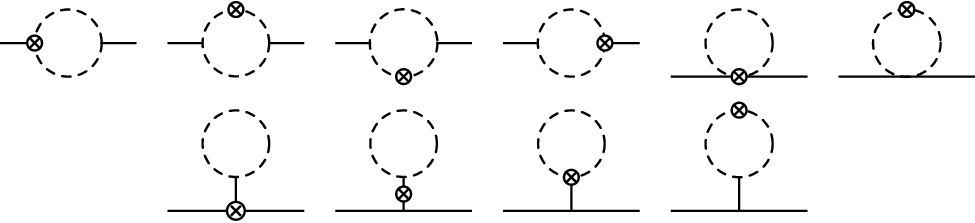,scale=0.8}
\caption{Topologies of one-loop diagrams contributing to the
  three-point vertex function of  the EMT operator and two Higgs boson
  fields. The crosses stand for the EMT insertions, solid lines refer
  to Higgs bosons, while dashed lines represent vector bosons, fermions, Higgs bosons, Goldstone bosons and Faddeev-Popov ghosts.}
\label{Higgs_EMT}
\end{center}
\vspace{-5mm}
\end{figure}

The one-loop topologies contributing to the three-point function are
shown in Fig.~\ref{Higgs_EMT}. By calculating these diagrams and
subsequently applying the LSZ formalism we extract the GFFs, whose
explicit expressions are given in appendix \ref{AppC}.
We find that $\theta_2 (q^2)$ is ultraviolet finite while $\theta_1
(q^2)$ diverges, and the corresponding divergence cannot be eliminated
by renormalization of the parameters in the electroweak Lagrangian, in
agreement with the results of Refs.~\cite{Callan:1970ze,Freedman:1974gs,Freedman:1974ze}. 
The (momentum-independent) divergent part of $\theta_1(q^2)$ is given by
\begin{equation}
\theta_1^{\rm div}= \frac{e^2 M_Z^2 \left(3 M_H^2+2 m_n^2-6 M_W^2-3 M_Z^2\right)}{24 \pi ^2
   (D-4) M_W^2\left(M_Z^2-M_W^2\right)} ,
\label{divpartD}
\end{equation}
where$M_Z$, $M_W$ and $m_n$ are the masses of the $Z$ and $W$ vector bosons, and $n$-th fermion, respectively,  while $e$ is the
electric charge. 
This divergence is canceled by the counter term generated by the
corresponding term of the Lagrangian $\sim R \phi^2$, where $R$ is the
scalar curvature \cite{Freedman:1974ze}. Such term appears in the EFT
Lagrangian that contains all local interactions compatible with the
symmetries of the electroweak theory and the general coordinate transformations.

\medskip

By considering localized one-particle states, the GFFs can be related
to the corresponding spatial distributions \cite{Panteleeva:2022uii}.
However, physical interpretation of the latter poses certain challenges. 
Problem occurs due to the fact that a superposition of eigenstates of the energy-momentum four-vector with different eigenvalues, which forms a localized state of the particle, is not an eigenstate of this operator. 
However, if the non-relativistic approximation is
  valid, the states can be defined by wave packets with the size much
larger than the Compton wavelength of the system (and much smaller
than characteristic radii of the system).  
For such wave packets, the integral over momentum
  eigenstates is governed 
by momenta much smaller than the mass of the system. Therefore, replacing the 
corresponding energies by the first term in the expansion $E=\sqrt{m^2+{\bf p}^2}=m+{\bf p}^2/(2 m)+\cdots $ provides a good approximation. 
That is, the packet is dominated by eigenstates of the energy with the
same eigenvalue $m$, and therefore it is also an (approximate)
eigenstate of the energy operator with the eigenvalue $m$.  
Thus,  $t^{00} ({\bf r}) $ can be interpreted in this case as the spatial distribution of 
mass, which in the static approximation coincides with the energy
distribution of the system.  
More details of the interpretation of the Fourier transforms of the gravitational form factors 
in the Breit frame in terms of  various spatial distributions can be found in Ref.~\cite{Polyakov:2018zvc}. 

In the case of the Higgs boson considered here, the static
approximation is by no means applicable as the slopes of the form
factors at zero momentum transfer,  $d \theta_i
/dq^2 \big|_{q^2 = 0}$, which give rise to the characteristic size of
the system, are smaller than the Compton wavelength squared of the Higgs boson.  
Therefore, we need to consider sharply localized states with the
size $R$ of the wave packet $\phi$ chosen much smaller than the Compton wave length. 
Such wave packet states are dominated by high momenta, and the spatial distribution given by  \cite{Panteleeva:2022uii}
 \be
\label{deft00} 
t^{00}({\bf r}) =N_{\phi,R} \, \int \frac{d^2 \hat n  \,d^3 {q}}{(2\pi)^3} \,
\theta_2 ( -{\bf q}_\perp^2 )    \, e^{-i {\bf q}\cdot {\bf  r}}
\ee
can be interpreted as energy distribution. Here, ${\bf q}_\perp = {\bf
  q} - {\bf q} \cdot \hat {\bf n} \hat {\bf n} $. 
Notice that while the overall normalization $N_{\phi,R}$ depends on
the form of the packet and diverges when the packet size is reduced,
the shape of the distribution does not depend on the details of the
(spherically symmetric)  wave packet used to prepare the system. Our
interpretation is that the normalization $N_{\phi,R}$ diverges in the
limit $R \to 0$ because one needs an increasing amount of energy to
reduce the size of the packet, while the shape of the distribution is
uniquely determined by the corresponding form factor and characterizes
the internal energy distribution of the system.

The energy distribution defined in  Eq.~(\ref{deft00}) leads to the mean-square energy-radius
\begin{equation}
r^2 =  4  \frac{d \theta_2 ( q^2 )}{dq^2} \bigg|_{q^2 = 0} \,.
\label{MSQER}
\end{equation}
From our result for $\theta_2(t )$ we obtain
\begin{eqnarray}
r^2 &=& -\frac{e^2 \left(6 m_n^2+M_H^2\right) M_Z^2
   B_0\left(M_H^2,m_n^2,m_n^2\right) m_n^4}{24 \pi ^2 M_H^4 \left(M_H^2-4
   m_n^2\right) M_W^2 \left(M_W^2-M_Z^2\right)}+\frac{e^2 \left(6
   m_n^2+M_H^2\right) M_Z^2 A_0\left(m_n^2\right) m_n^2}{24 \pi ^2 M_H^4
   \left(M_H^2-4 m_n^2\right) M_W^2 \left(M_W^2-M_Z^2\right)} \nonumber\\   
&+& \frac{e^2 M_Z^2 A_0\left(M_H^2\right)}{24 \pi
   ^2 M_H^2 M_W^2 \left(M_W^2-M_Z^2\right)}+\frac{e^2 \left(5 M_H^6-29
   M_W^2 M_H^4+24 M_W^4 M_H^2+36 M_W^6\right) M_Z^2
   A_0\left(M_W^2\right)}{24 \pi ^2 M_H^4 M_W^2 \left(M_H^2-4
   M_W^2\right){}^2 \left(M_W^2-M_Z^2\right)} \nonumber\\
   &+& \frac{e^2 M_Z^2 \left(5
   M_H^6-29 M_Z^2 M_H^4+24 M_Z^4 M_H^2+36 M_Z^6\right)
   A_0\left(M_Z^2\right)}{48 \pi ^2 M_H^4 M_W^2 \left(M_H^2-4
   M_Z^2\right){}^2 \left(M_W^2-M_Z^2\right)}-\frac{e^2 M_Z^2
   B_0\left(M_H^2,M_H^2,M_H^2\right)}{24 \pi ^2 M_W^2
   \left(M_W^2-M_Z^2\right)} 
   \nonumber\\
   &-& \frac{e^2 \left(5 M_H^6-29 M_W^2 M_H^4+24
   M_W^4 M_H^2+36 M_W^6\right) M_Z^2 B_0\left(M_H^2,M_W^2,M_W^2\right)}{24
   \pi ^2 M_H^4 \left(M_H^2-4 M_W^2\right){}^2
   \left(M_W^2-M_Z^2\right)} 
    \nonumber\\
    &-& \frac{e^2 M_Z^4 \left(5 M_H^6-29 M_Z^2
   M_H^4+24 M_Z^4 M_H^2+36 M_Z^6\right)
   B_0\left(M_H^2,M_Z^2,M_Z^2\right)}{48 \pi ^2 M_H^4 M_W^2 \left(M_H^2-4
   M_Z^2\right){}^2 \left(M_W^2-M_Z^2\right)}+\frac{e^2 \left(M_H^6-24
   m_n^6\right) M_Z^2}{96 \pi ^2 M_H^4 \left(M_H^2-4 m_n^2\right) M_W^2
   \left(M_W^2-M_Z^2\right)} \nonumber\\
   &-& \frac{e^2
   M_Z^2 }{96 \pi ^2 M_H^4 M_W^2 \left(M_H^2-4
   M_W^2\right){}^2 \left(M_H^2-4 M_Z^2\right){}^2
   \left(M_W^2-M_Z^2\right)} 
   \Bigl[
   144 \left(M_H^2-4 M_Z^2\right){}^2 M_W^8 \nonumber\\
   &+& 72 \left(M_H^3-4
   M_H M_Z^2\right){}^2 M_W^6+16 \left(5 M_H^8-41 M_Z^2 M_H^6+72 M_Z^4
   M_H^4+36 M_Z^6 M_H^2+72 M_Z^8\right) M_W^4 
   \nonumber\\
   &-&
   2 M_H^2 \left(25 M_H^8-204
   M_Z^2 M_H^6+368 M_Z^4 M_H^4+144 M_Z^6 M_H^2+288 M_Z^8\right) M_W^2+M_H^4
   \left(6 M_H^8-49 M_Z^2 M_H^6 \right. \nonumber\\
   &+& \left. 88 M_Z^4 M_H^4+36 M_Z^6 M_H^2+72
   M_Z^8\right)
   \Bigr]
    \,.
   \label{massradius}
\end{eqnarray}
Substituting numerical values of various parameters from Ref.~\cite{ParticleDataGroup:2024cfk} we estimate:
\begin{equation}
r^2 =  1.44\times {10}^{-6}+(5.49\times {10}^{-8} + 1.10\times {10}^{-12} \, i) \ {\rm GeV}^{-2}\,,
\label{MSQERNUM}
\end{equation}
where the number in the brackets refers to the contribution of
fermions. Notice that this contribution also contains an imaginary
part as the Higgs particle is unstable,  decaying in fermion-anti-fermion pairs.

Our expression of the one-loop correction to the D-term of the Higgs boson has the form:
\begin{eqnarray}
D &=& -\frac{e^2 m_n^2 M_Z^2 B_0\left(M_H^2,m_n^2,m_n^2\right)}{48 \pi ^2 M_W^2
   \left(M_W^2-M_Z^2\right)} -\frac{e^2 M_H^2 M_Z^2
   B_0\left(M_H^2,M_H^2,M_H^2\right)}{32 \pi ^2 M_W^2
   \left(M_W^2-M_Z^2\right)} + \frac{e^2 M_Z^2 A_0\left(m_n^2\right)}{16 \pi
   ^2 M_W^2 \left(M_W^2-M_Z^2\right)} \nonumber\\
   &+& \frac{e^2 M_Z^2 \left(4 M_H^2 M_W^2-M_H^4+12
   M_W^4\right) B_0\left(M_H^2,M_W^2,M_W^2\right)}{48 \pi ^2 M_W^2 \left(4
   M_W^2-M_H^2\right) \left(M_W^2-M_Z^2\right)}+\frac{e^2 M_Z^2 \left(4
   M_H^2 M_Z^2-M_H^4+12 M_Z^4\right) B_0\left(M_H^2,M_Z^2,M_Z^2\right)}{96
   \pi ^2 M_W^2 \left(4 M_Z^2-M_H^2\right)
   \left(M_W^2-M_Z^2\right)} \nonumber\\
   &+& \frac{e^2 M_Z^2 A_0\left(M_H^2\right)}{16 \pi
   ^2 M_W^2 \left(M_W^2-M_Z^2\right)}-\frac{e^2 M_Z^2 A_0\left(M_W^2\right)
   \left(6 M_W^2-M_H^2\right)}{8 \pi ^2 M_W^2 \left(4 M_W^2-M_H^2\right)
   \left(M_W^2-M_Z^2\right)} 
   + \frac{e^2 M_Z^2 A_0\left(M_Z^2\right)
   \left(M_H^2-6 M_Z^2\right)}{16 \pi ^2 M_W^2 \left(4 M_Z^2-M_H^2\right)
   \left(M_W^2-M_Z^2\right)}\nonumber\\
   &-& 
   \frac{e^2 M_Z^2 }{48 \pi ^2
   M_W^2 \left(4 M_W^2-M_H^2\right) \left(4 M_Z^2-M_H^2\right)
   \left(M_W^2-M_Z^2\right)} 
   \Bigl[
   -m_n^2 \left(4
   M_W^2-M_H^2\right) \left(M_H^2-4 M_Z^2\right)+4 M_W^4 \left(4
   M_Z^2-M_H^2\right) \nonumber\\
   &+& 2 M_W^2 \left(30 M_H^2 M_Z^2-7 M_H^4+4
   M_Z^4\right)-13 M_H^4 M_Z^2-2 M_H^2 M_Z^4+3 M_H^6
   \Bigr] 
   \,.
\label{Dterm}
\end{eqnarray}
As mentioned above, the $D$-term contains ultraviolet divergence which
is cancelled by corresponding parameter of the effective Lagrangian.

\section{Summary}
\label{conclusions}

In the current work we have calculated the electroweak correction to
the matrix element of the EMT of the Higgs boson to one loop and extracted the corresponding GFFs. 
We found that the $\theta_2(q^2)$ GFF is ultraviolet finite, while
the $\theta_1 (q^2)$ diverges, in agreement with Refs.~\cite{Callan:1970ze,Freedman:1974gs,Freedman:1974ze}. 
This divergence cannot be cancelled by counter terms generated by the
Lagrangian of the electroweak theory minimally coupled to the
gravitational field. However the corresponding dimension-four operator
is present in the standard-model EFT Lagrangian that contains all local interactions compatible with underlying symmetries \cite{Freedman:1974ze}.
By considering matrix elements of the EMT operator for localized wave
packet states, the GFFs can be related to corresponding spatial distributions \cite{Panteleeva:2022uii}.
Our expressions for both form factors $\theta_1(q^2)$ and $\theta_2
(q^2)$ have non-trivial dependence on the momentum transfer squared.
The characteristic radius of the energy distribution of the Higgs boson, defined
by the slope of the form factor $\theta_2(q^2)$ at vanishing
momentum transfer, is smaller than the Compton wave length of the
Higgs boson. Therefore, it is not appropriate to consider the
Breit-frame densities corresponding to the static approximation.    
On the other hand, the gravitational interaction probes the energy
distribution of the Higgs boson (and reveals a non-zero energy-radius) 
through the matrix element of the EMT operator between sharply
localized states with the wave packet size chosen to be much smaller than
the Compton wavelength of the Higgs boson. 
Clearly, it is not feasible to measure  gravitational scattering off
the Higgs boson in experiment, and it is also unfeasible to localize
it at distances smaller than its Compton radius. Still, our
theoretical investigation demonstrates that an electrically neutral
elementary particle reveals internal structure with a non-vanishing
spatial extension when being probed by the gravitational interaction.   
One might speculate that in analogy with charged particles, the
non-zero radius emerges due to the weak field of massive vector
particles, however we find that similar in size contributions to the
radius are generated by diagrams with fermion loops and
self-interactions  of the scalar field.

\acknowledgements

This work was supported in part 
by the MKW NRW under the funding code NW21-024-A, by the Georgian
Shota Rustaveli National 
Science Foundation (Grant No. FR-23-856), 
by the European Research Council (ERC) under the European Union's
Horizon 2020 research and innovation programme (grant agreement
No. 885150)
and by the EU Horizon 2020 research and
innovation programme (STRONG-2020, grant agreement No. 824093).

\appendix
\section{Definition of loop integrals}
\label{AppA}

One-loop integrals are defined as follows:
\begin{eqnarray}
 \text{A}_0(m^2)&=&\frac{(2\pi)^{4-D}}{i\pi^2} \int\frac{d^D k}{k^2-m^2+i \epsilon}\,,\nonumber\\
 \text{B}_0(p^2,m_1^2,m_2^2)&=&\frac{(2\pi)^{4-D}}{i\pi^2} \int\frac{d^Dk}{[k^2-m_1^2+i \epsilon] [(p+k)^2-m_2^2+i \epsilon]}\,, 
 \nonumber\\
C_0(p_1^2,p_2^2,p^2_{12},m_1^2,m_2^2,m_3^2)&=&\frac{(2\pi)^{4-n}}{i\pi^2} \int\frac{d^nk}{[k^2-m_1^2+i \epsilon] [(p_1+k)^2-m_2^2+i \epsilon] [(p_1+p_2+k)^2-m_ 3^2+i \epsilon]}\,,
\label{ints}
\end{eqnarray}
where $p_{12}=p_1+p_2$ and $D$ is the spacetime dimension.
For tensor loop integrals, we apply the reduction formulae of Ref.~\cite{Denner:2005nn},  
while for the expansion of the scalar integrals in Eq.~(\ref{ints}) in terms of kinematical invariants we use Ref.~\cite{Devaraj:1997es}.

\section{Contributions to the pole position and the residue of the
  Higgs propagator}
\label{AppB}

One-loop contributions to the pole position and the residue of the Higgs-boson propagator are given by
\begin{eqnarray}
\delta z_1 & = & \frac{ - e^2 M_Z^2 }{128 \pi ^2 M_W^2
   \left(M_W^2-M_Z^2\right)}  
   \left(  \left(-8 (D-1) M_W^4+8 M_H^2 M_W^2-2 M_H^4\right)
   \text{B}_0\left(M_H^2,M_W^2,M_W^2\right) \right. \nonumber\\
   &+& \left. \left(-4 (D-1) M_Z^4+4 M_H^2
   M_Z^2-M_H^4\right) \text{B}_0\left(M_H^2,M_Z^2,M_Z^2\right)-4 m_n^2
   \left(M_H^2-4 m_n^2\right) \text{B}_0\left(M_H^2,m_n^2,m_n^2\right) \right. \nonumber\\
   &-& \left. 9 M_H^4
   \text{B}_0\left(M_H^2,M_H^2,M_H^2\right)+4 \text{A}_0\left(M_W^2\right)
   \left(2 (D-1) M_W^2+M_H^2\right)+2 \text{A}_0\left(M_Z^2\right) \left(2
   (D-1) M_Z^2+M_H^2\right) \right. \nonumber\\
   &+& \left. 6 M_H^2 \text{A}_0\left(M_H^2\right)-16 m_n^2
   \text{A}_0\left(m_n^2\right)
   \right)\,,\nonumber\\
   \delta Z_1 &=& -\frac{e^2 M_Z^2}{128 \pi ^2 M_W^2 \left(M_W^2-M_Z^2\right)} 
   \Biggl[ 
 -\frac{4 \left(4 (D-1) M_W^6-5 M_H^4 M_W^2+4 M_H^2 M_W^4+M_H^6\right)
   \text{B}_0\left(M_H^2,M_W^2,M_W^2\right)}{M_H^4-4 M_H^2 M_W^2}  \nonumber\\
   &-&  \frac{2
   \left(4 (D-1) M_Z^6-5 M_H^4 M_Z^2+4 M_H^2 M_Z^4+M_H^6\right)
   \text{B}_0\left(M_H^2,M_Z^2,M_Z^2\right)}{M_H^4-4 M_H^2
   M_Z^2} - \left(\frac{8 m_n^4}{M_H^2} + 4 m_n^2\right)
   \text{B}_0\left(M_H^2,m_n^2,m_n^2\right)  \nonumber\\
   &+& 4 M_H^2 \text{B}_0\left(M_H^2,M_W^2
   \alpha _W,M_W^2 \alpha _W\right)+2 M_H^2 \text{B}_0\left(M_H^2,M_Z^2 \alpha
   _Z,M_Z^2 \alpha _Z\right)+6 M_H^2
   \text{B}_0\left(M_H^2,M_H^2,M_H^2\right) \nonumber\\ 
   &+& \frac{8
   \text{A}_0\left(M_W^2\right) \left(2 (D-1) M_W^4-4 M_H^2
   M_W^2+M_H^4\right)}{M_H^4-4 M_H^2 M_W^2}+\frac{4
   \text{A}_0\left(M_Z^2\right) \left(2 (D-1) M_Z^4-4 M_H^2
   M_Z^2+M_H^4\right)}{M_H^4-4 M_H^2 M_Z^2} \nonumber\\
   &+& \frac{8 m_n^2
   \text{A}_0\left(m_n^2\right)}{M_H^2}-6 \text{A}_0\left(M_H^2\right)-4
   \text{A}_0\left(M_W^2 \alpha _W\right)-2 \text{A}_0\left(M_Z^2 \alpha
   _Z\right) \nonumber\\
   &+& \frac{1}{M_H^2 \left(M_H^2-4 M_W^2\right) \left(M_H^2-4 M_Z^2\right)} \left( 8 (D+1) M_H^2 M_W^4 \left(M_H^2-4 M_Z^2\right)+16 (D-1)
   M_W^6 \left(4 M_Z^2-M_H^2\right) \right. \nonumber\\
   &+&  \left. 4 M_W^2 \left(-4 (D+1) M_H^2 M_Z^4+8 (D-1)
   M_Z^6+30 M_H^4 M_Z^2-7 M_H^6\right)+4 (D+1) M_H^4 M_Z^4-8 (D-1) M_H^2
   M_Z^6 \right. \nonumber\\ 
   &+& \left. 4 M_H^2 m_n^2 \left(M_H^2-4 M_W^2\right) \left(M_H^2-4 M_Z^2\right)+8
   m_n^4 \left(4 M_W^2-M_H^2\right) \left(M_H^2-4 M_Z^2\right)-26 M_H^6 M_Z^2+6
   M_H^8
  \right) \Biggr] \,.
\label{HSE}
\end{eqnarray}

\section{Gravitational form factors}
\label{AppC}

One-loop contributions to the GFFs of the Higgs boson have the form
\begin{eqnarray}
\theta_1 (t) &= & \frac{9 e^2 \left(12 M_H^4+4 (D-3) t M_H^2-(D-2) t^2\right) M_Z^2
    M_H^4 C_0\left(M_H^2,M_H^2,t,M_H^2,M_H^2,M_H^2\right) }{64 (D-2) \pi ^2 t
   \left(4 M_H^2-t\right) M_W^2 \left(M_W^2-M_Z^2\right)} \nonumber\\
   &-& \frac{3 e^2
   \left(8 (D-5) M_H^6+4 \left(2 D^2+3 D-5\right) t M_H^4+2 \left(D^2-12
   D+14\right) t^2 M_H^2-(D-2)^2 t^3\right) M_Z^2
    M_H^2 B_0\left(t,M_H^2,M_H^2\right) }{128 (D-2) (D-1) \pi ^2 t
   \left(M_H^2-t\right) \left(4 M_H^2-t\right) M_W^2
   \left(M_W^2-M_Z^2\right)}\nonumber\\
   &+& \frac{3 e^2 \left(-12 (D-3) M_H^4-(D-2) t
   M_H^2+(D-2) t^2\right) M_Z^2  M_H^2 B_0\left(M_H^2,M_H^2,M_H^2\right) }{64
   (D-2) \pi ^2 t \left(4 M_H^2-t\right) M_W^2
   \left(M_W^2-M_Z^2\right)}\nonumber\\
   &+& \frac{3 e^2 \left(2 (D-2) M_H^4+(2 D-3) t
   M_H^2-(D-1) t^2\right) M_Z^2 A_0\left(M_H^2\right)}{64 (D-1) \pi ^2 t
   \left(M_H^2-t\right) M_W^2 \left(M_W^2-M_Z^2\right)} \nonumber\\
   &+& \frac{e^2 m_n^2 M_Z^2
   B_0\left(t,m_n^2,m_n^2\right)}{16 (D-2) (D-1) \pi ^2 t
   \left(M_H^2-t\right) \left(4 M_H^2-t\right) M_W^2
   \left(M_W^2-M_Z^2\right)}
   \biggl[
   4 (D-1) M_H^6-8 (2 D-3) t M_H^4 \nonumber\\
   &+& (3 D-1) t^2 M_H^2-t^3+4 m_n^2
   \left(4 (D-3) M_H^4+2 (4 D-5) t M_H^2+(4-3 D) t^2\right)
   \biggr] \nonumber\\
   &+& \frac{e^2 M_Z^2 B_0\left(t,M_W^2,M_W^2\right)}{64 (D-2) (D-1) \pi ^2
   t \left(M_H^2-t\right) \left(4 M_H^2-t\right) M_W^2
   \left(M_W^2-M_Z^2\right)} 
   \biggl[ 
   8 (D-1) M_H^8-4 \left(2 D^2-5
   D+5\right) t M_H^6 \nonumber\\
   &-&2 \left(D^2-6 D+6\right) t^2 M_H^4+(D-2)^2 t^3
   M_H^2-8 \left(4 \left(D^2-4 D+3\right) M_H^4+2 \left(4 D^2-9 D+5\right)
   t M_H^2 \right. \nonumber\\ 
   &+& \left. \left(-3 D^2+7 D-4\right) t^2\right) M_W^4-2 \left(16 (2 D-3)
   M_H^6+4 \left(2 D^3-18 D^2+31 D-16\right) t M_H^4 \right. \nonumber\\ 
   &+& \left. 2 \left(D^3-9 D^2+35
   D-26\right) t^2 M_H^2-\left(D^3-9 D^2+28 D-20\right) t^3\right)
   M_W^2
   \biggr] 
   \nonumber\\
   &+& \frac{e^2 M_Z^2 B_0\left(t,M_Z^2,M_Z^2\right)}{128 (D-2) (D-1) \pi ^2 t
   \left(M_H^2-t\right) \left(4 M_H^2-t\right) M_W^2
   \left(M_W^2-M_Z^2\right)}
   \biggl[
   8 (D-1) M_H^8-4 \left(2
   D^2-5 D+5\right) t M_H^6  \nonumber\\
   &-& 2 \left(D^2-6 D+6\right) t^2 M_H^4+(D-2)^2 t^3
   M_H^2-8 \left(4 \left(D^2-4 D+3\right) M_H^4+2 \left(4 D^2-9 D+5\right)
   t M_H^2 \right. \nonumber\\ 
   &+& \left. \left(-3 D^2+7 D-4\right) t^2\right) M_Z^4-2 \left(16 (2 D-3)
   M_H^6+4 \left(2 D^3-18 D^2+31 D-16\right) t M_H^4 \right. \nonumber\\
   &+& \left. 2 \left(D^3-9 D^2+35
   D-26\right) t^2 M_H^2-\left(D^3-9 D^2+28 D-20\right) t^3\right)
   M_Z^2
   \biggr] 
   \nonumber\\
   &+& \frac{e^2 m_n^2 \left(-2 M_H^6+t M_H^4+8 m_n^4
   \left(t-4 M_H^2\right)+m_n^2 \left(16 M_H^4-6 t M_H^2+t^2\right)\right)
   M_Z^2 C_0\left(M_H^2,M_H^2,t,m_n^2,m_n^2,m_n^2\right)}{8 (D-2) \pi ^2 t
   \left(4 M_H^2-t\right) M_W^2 \left(M_W^2-M_Z^2\right)}\nonumber\\
   &+& \frac{e^2 M_Z^2
   C_0\left(M_H^2,M_H^2,t,M_W^2,M_W^2,M_W^2\right)}{32 (D-2) \pi ^2 t
   \left(4 M_H^2-t\right) M_W^2 \left(M_W^2-M_Z^2\right)} 
   \biggl[
   -4 M_H^8+4 (D-2) t M_H^6-(D-2) t^2 M_H^4 \nonumber\\
   &+& 16 (D-1) \left(4
   M_H^2-t\right) M_W^6-4 \left(4 (D+3) M_H^4-4 \left(D^2-7 D+15\right) t
   M_H^2+\left(D^2-7 D+14\right) t^2\right) M_W^4 \nonumber\\
   &+& 2 \left(16 M_H^6+(6-8 D)
   t M_H^4+6 (D-2) t^2 M_H^2-(D-2) t^3\right) M_W^2
   \biggr] 
   \nonumber\\
   &+& \frac{e^2 M_Z^2 C_0\left(M_H^2,M_H^2,t,M_Z^2,M_Z^2,M_Z^2\right)}{64 (D-2) \pi ^2 t
   \left(4 M_H^2-t\right) M_W^2 \left(M_W^2-M_Z^2\right)}
   \biggl[
   -4 M_H^8+4 (D-2) t M_H^6-(D-2) t^2 M_H^4 \nonumber\\ 
   &+&16 (D-1) \left(4
   M_H^2-t\right) M_Z^6-4 \left(4 (D+3) M_H^4-4 \left(D^2-7 D+15\right) t
   M_H^2+\left(D^2-7 D+14\right) t^2\right) M_Z^4 \nonumber\\
   &+& 2 \left(16 M_H^6+(6-8 D)
   t M_H^4+6 (D-2) t^2 M_H^2-(D-2) t^3\right) M_Z^2
   \biggr]
    \nonumber\\
   &-& \frac{e^2 M_Z^2 }{64 \pi ^2 M_W^2 \left(4
   M_W^2-M_H^2\right) \left(M_W^2-M_Z^2\right) \left(4 M_Z^2-M_H^2\right)
   M_H^2}
   \biggl[
   3 M_H^8-13 M_Z^2 M_H^6+2 (D+1) M_Z^4 M_H^4 \nonumber\\
   &-& 4 (D-1) M_Z^6 M_H^2+4
   (D+1) M_W^4 \left(M_H^2-4 M_Z^2\right) M_H^2+2 m_n^2 \left(M_H^2-4
   M_W^2\right) \left(M_H^2-4 M_Z^2\right) M_H^2 \nonumber\\ 
   &+& 4 m_n^4 \left(4
   M_W^2-M_H^2\right) \left(M_H^2-4 M_Z^2\right)+8 (D-1) M_W^6 \left(4
   M_Z^2-M_H^2\right)\nonumber\\
   &+&2 M_W^2 \left(-7 M_H^6+30 M_Z^2 M_H^4-4 (D+1) M_Z^4
   M_H^2+8 (D-1) M_Z^6\right)
   \biggr]
   \nonumber\\
   &+& \frac{e^2 m_n^2 \left(-4 (D-2) M_H^4+(5-3 D) t M_H^2+(D-1)
   t^2\right) M_Z^2 A_0\left(m_n^2\right)}{16 (D-1) \pi ^2 t
   \left(M_H^2-t\right) M_W^2 \left(M_W^2-M_Z^2\right) M_H^2} \nonumber
   \end{eqnarray}
   
    \begin{eqnarray} 
 &+& \frac{e^2 M_Z^2
   A_0\left(M_W^2\right)}{32 (D-1) \pi ^2 t \left(M_H^2-t\right) M_W^2
   \left(4 M_W^2-M_H^2\right) \left(M_W^2-M_Z^2\right) M_H^2}
   \biggl[
    -2 (D-2) M_H^8+t M_H^6-(D-1) t^2 M_H^4 \nonumber\\
    &+& 4 \left(4 \left(D^2-3
   D+2\right) M_H^4+\left(3 D^2-8 D+5\right) t M_H^2-(D-1)^2 t^2\right)
   M_W^4-2 \left(2 \left(D^2-5 D+6\right) M_H^6 \right. \nonumber\\
   &+& \left. \left(D^2-3 D+4\right) t
   M_H^4-2 (D-1) t^2 M_H^2\right) M_W^2
   \biggr] 
   \nonumber\\
   &+& \frac{e^2
   M_Z^2 A_0\left(M_Z^2\right)
   }{64
   (D-1) \pi ^2 t \left(M_H^2-t\right) M_W^2 \left(M_W^2-M_Z^2\right)
   \left(4 M_Z^2-M_H^2\right) M_H^2} 
   \biggl[ 
   -2 (D-2) M_H^8+t M_H^6-(D-1) t^2 M_H^4  \nonumber\\
   &+& 4 \left(4 \left(D^2-3
   D+2\right) M_H^4+\left(3 D^2-8 D+5\right) t M_H^2-(D-1)^2 t^2\right)
   M_Z^4-2 \left(2 \left(D^2-5 D+6\right) M_H^6 \right. \nonumber\\
   &+& \left. \left(D^2-3 D+4\right) t
   M_H^4-2 (D-1) t^2 M_H^2\right) M_Z^2
   \biggr]
   \nonumber\\
   &-& \frac{e^2 \left(\left(2 (D-2) t^2-32
   (D-3) M_H^4\right) m_n^4+M_H^2 \left(2 M_H^2-t\right) \left(4 (D-3)
   M_H^2-(D-2) t\right) m_n^2\right) M_Z^2
   B_0\left(M_H^2,m_n^2,m_n^2\right)}{32 (D-2) \pi ^2 t \left(4
   M_H^2-t\right) M_W^2 \left(M_W^2-M_Z^2\right) M_H^2} \nonumber\\
   &-& \frac{e^2 M_Z^2 B_0\left(M_H^2,M_W^2,M_W^2\right)}{32
   (D-2) \pi ^2 t \left(4 M_H^2-t\right) M_W^2 \left(4 M_W^2-M_H^2\right)
   \left(M_W^2-M_Z^2\right) M_H^2} 
   \biggl[
    \left((D-2) t-4 (D-3) M_H^2\right) M_H^8  \nonumber\\
    &+& 4 \left(16 \left(D^2-4
   D+3\right) M_H^4-\left(D^2-3 D+2\right) t^2\right) M_W^6-4 \left(4
   \left(D^2-9\right) M_H^6-\left(D^2+13 D-46\right) t M_H^4 \right.  \nonumber\\ 
   &+& \left. 3 (D-2) t^2
   M_H^2\right) M_W^4+\left(32 (D-3) M_H^8+4 (14-5 D) t M_H^6+3 (D-2) t^2
   M_H^4\right) M_W^2 
   \biggr] 
   \nonumber\\
   &-& \frac{e^2 M_Z^2  B_0\left(M_H^2,M_Z^2,M_Z^2\right)  }{64 (D-2)
   \pi ^2 t \left(4 M_H^2-t\right) M_W^2 \left(M_W^2-M_Z^2\right) \left(4
   M_Z^2-M_H^2\right) M_H^2} 
   \biggl[ 
   \left((D-2) t-4
   (D-3) M_H^2\right) M_H^8 \nonumber\\
   &+& 4 \left(16 \left(D^2-4 D+3\right)
   M_H^4-\left(D^2-3 D+2\right) t^2\right) M_Z^6-4 \left(4
   \left(D^2-9\right) M_H^6-\left(D^2+13 D-46\right) t M_H^4 \right. \nonumber\\
   &+& \left. 3 (D-2) t^2
   M_H^2\right) M_Z^4+\left(32 (D-3) M_H^8+4 (14-5 D) t M_H^6+3 (D-2) t^2
   M_H^4\right) M_Z^2
   \biggr] 
   \,.
\label{theta1Explicit}
\end{eqnarray}

\begin{eqnarray}
\theta_2 (t) &= &   \frac{9 e^2 \left(4
   (D-5) M_H^4-4 (D-3) t M_H^2+(D-2) t^2\right) M_Z^2
   C_0\left(M_H^2,M_H^2,t,M_H^2,M_H^2,M_H^2\right) M_H^4}{64 (D-2) \pi ^2
   \left(t-4 M_H^2\right){}^2 M_W^2 \left(M_W^2-M_Z^2\right)}\nonumber\\
   &+& \frac{3 e^2
   \left(4 (2 D-7) M_H^4-(D-2) t M_H^2-(D-2) t^2\right) M_Z^2
   B_0\left(M_H^2,M_H^2,M_H^2\right) M_H^2}{64 (D-2) \pi ^2 \left(t-4
   M_H^2\right){}^2 M_W^2 \left(M_W^2-M_Z^2\right)}+\frac{3 e^2 M_Z^2
   A_0\left(M_H^2\right)}{64 \pi ^2 M_W^2
   \left(M_W^2-M_Z^2\right)} \nonumber\\
   &-& \frac{e^2 m_n^2 \left(4 (D-1) M_H^4-4 t
   M_H^2+t^2-8 (D-1) m_n^2 \left(2 M_H^2-t\right)\right) M_Z^2
   B_0\left(t,m_n^2,m_n^2\right)}{16 (D-2) \pi ^2 \left(t-4
   M_H^2\right){}^2 M_W^2 \left(M_W^2-M_Z^2\right)} \nonumber\\
   &-& \frac{e^2 M_Z^2
   B_0\left(t,M_W^2,M_W^2\right)}{16 (D-2) \pi ^2 \left(t-4
   M_H^2\right){}^2 M_W^2 \left(M_W^2-M_Z^2\right)} 
   \biggl[
   (D-1)
   \left(2 M_H^2-t\right) M_H^4+4 (D-1)^2 \left(2 M_H^2-t\right) M_W^4 \nonumber\\
   &-& 4
   \left(2 (D-1) M_H^4+(D-5) t M_H^2+t^2\right) M_W^2
   \biggr] -\frac{9 (D-1) e^2 \left(2 M_H^2-t\right) M_Z^2
   B_0\left(t,M_H^2,M_H^2\right) M_H^4}{32 (D-2) \pi ^2 \left(t-4
   M_H^2\right){}^2 M_W^2 \left(M_W^2-M_Z^2\right)}
    \nonumber\\
   &-& \frac{e^2 M_Z^2 B_0\left(t,M_Z^2,M_Z^2\right)}{32 (D-2) \pi ^2 \left(t-4
   M_H^2\right){}^2 M_W^2 \left(M_W^2-M_Z^2\right)}
   \biggl[
   (D-1) \left(2 M_H^2-t\right) M_H^4+4 (D-1)^2 \left(2
   M_H^2-t\right) M_Z^4 \nonumber\\
   &-& 4 \left(2 (D-1) M_H^4+(D-5) t M_H^2+t^2\right)
   M_Z^2
   \biggr] 
    \nonumber\\
   &+& \frac{e^2 m_n^2 M_Z^2
   C_0\left(M_H^2,M_H^2,t,m_n^2,m_n^2,m_n^2\right)}{8 (D-2) \pi ^2
   \left(t-4 M_H^2\right){}^2 M_W^2 \left(M_W^2-M_Z^2\right)}
   \biggl[
   8 \left(4 M_H^2-t\right) m_n^4+\left(t \left(3 t-10 M_H^2\right)-2 D
   \left(t-2 M_H^2\right){}^2\right) m_n^2  \nonumber\\
   &+& (D-1) M_H^4 \left(2
   M_H^2-t\right)
   \biggr]
    \nonumber\\
   &+& \frac{e^2 M_Z^2
   C_0\left(M_H^2,M_H^2,t,M_W^2,M_W^2,M_W^2\right)}{32 (D-2) \pi ^2
   \left(t-4 M_H^2\right){}^2 M_W^2 \left(M_W^2-M_Z^2\right)}
   \biggl[
   4 (D-1) M_H^8-4 (D-2) t M_H^6+(D-2) t^2 M_H^4  \nonumber\\
   &-& 16 (D-1) \left(4
   M_H^2-t\right) M_W^6+4 \left(4 \left(D^2-2 D+5\right) M_H^4-4
   \left(D^2-3 D+7\right) t M_H^2+\left(D^2-3 D+6\right) t^2\right) M_W^4  \nonumber\\
   &-& 2
   \left(2 t \left(11 M_H^4-6 t M_H^2+t^2\right)+D \left(8 M_H^6-16 t
   M_H^4+6 t^2 M_H^2-t^3\right)\right) M_W^2
   \biggr]
   \nonumber\\
   &+& \frac{e^2 M_Z^2    C_0\left(M_H^2,M_H^2,t,M_Z^2,M_Z^2,M_Z^2\right)}{64 (D-2) \pi ^2
   \left(t-4 M_H^2\right){}^2 M_W^2 \left(M_W^2-M_Z^2\right)}
   \biggl[ 
   4 (D-1) M_H^8-4 (D-2) t M_H^6+(D-2) t^2 M_H^4 \nonumber\\
   &-& 16 (D-1)
   \left(4 M_H^2-t\right) M_Z^6+4 \left(4 \left(D^2-2 D+5\right) M_H^4-4
   \left(D^2-3 D+7\right) t M_H^2+\left(D^2-3 D+6\right) t^2\right) M_Z^4 \nonumber\\
   &-& 2
   \left(2 t \left(11 M_H^4-6 t M_H^2+t^2\right)+D \left(8 M_H^6-16 t
   M_H^4+6 t^2 M_H^2-t^3\right)\right) M_Z^2
   \biggr]
 \nonumber\\
   &-& \frac{e^2
   M_Z^2 }{64 \pi ^2 M_W^2 \left(4
   M_W^2-M_H^2\right) \left(M_W^2-M_Z^2\right) \left(4 M_Z^2-M_H^2\right)
   M_H^2} 
   \biggl[
   3 M_H^8-13 M_Z^2 M_H^6+2 (D+1) M_Z^4 M_H^4 \nonumber\\
   &- & 4 (D-1) M_Z^6
   M_H^2+4 (D+1) M_W^4 \left(M_H^2-4 M_Z^2\right) M_H^2+2 m_n^2
   \left(M_H^2-4 M_W^2\right) \left(M_H^2-4 M_Z^2\right) M_H^2 \nonumber\\
   &+& 4 m_n^4
   \left(4 M_W^2-M_H^2\right) \left(M_H^2-4 M_Z^2\right)+8 (D-1) M_W^6
   \left(4 M_Z^2-M_H^2\right) \nonumber\\
   &+& 2 M_W^2 \left(-7 M_H^6+30 M_Z^2 M_H^4-4 (D+1)
   M_Z^4 M_H^2+8 (D-1) M_Z^6\right)
   \biggr] \nonumber\\
   &-& \frac{e^2 m_n^2 M_Z^2 A_0\left(m_n^2\right)}{16 \pi ^2 M_W^2
   \left(M_W^2-M_Z^2\right) M_H^2}+\frac{e^2 \left(M_H^4-4 M_W^2 M_H^2+4
   (D-1) M_W^4\right) M_Z^2 A_0\left(M_W^2\right)}{32 \pi ^2 M_W^2 \left(4
   M_W^2-M_H^2\right) \left(M_W^2-M_Z^2\right) M_H^2} \nonumber\\
   &+& \frac{e^2 m_n^2 M_Z^2 B_0\left(M_H^2,m_n^2,m_n^2\right)}{32 (D-2) \pi
   ^2 \left(t-4 M_H^2\right){}^2 M_W^2 \left(M_W^2-M_Z^2\right)
   M_H^2} 
   \biggl[
   8 (2 D-5) M_H^6+2 (8-3 D) t M_H^4+(D-2) t^2
   M_H^2 \nonumber\\
   &+& m_n^2 \left(-32 (D-3) M_H^4+8 (D-2) t M_H^2+2 (D-2)
   t^2\right)
   \biggr]  \nonumber\\
   &+& \frac{e^2 M_Z^2 B_0\left(M_H^2,M_W^2,M_W^2\right)}{32
   (D-2) \pi ^2 \left(t-4 M_H^2\right){}^2 M_W^2 \left(4 M_W^2-M_H^2\right)
   \left(M_W^2-M_Z^2\right) M_H^2} \biggl[
   \left((20-8 D) M_H^2+3 (D-2) t\right) M_H^8 \nonumber\\
   &+& 4
   \left(16 \left(D^2-4 D+3\right) M_H^4-4 \left(D^2-3 D+2\right) t
   M_H^2-\left(D^2-3 D+2\right) t^2\right) M_W^6 -4 \left(4 \left(2 D^2-3
   D-7\right) M_H^6 \right.  \nonumber\\
   &+& \left. \left(-3 D^2+5 D+18\right) t M_H^4+3 (D-2) t^2
   M_H^2\right) M_W^4+\left(16 (3 D-8) M_H^8-16 (D-3) t M_H^6+3 (D-2) t^2
   M_H^4\right) M_W^2
   \biggr]  \nonumber\\
   &+& \frac{e^2 M_Z^2  B_0\left(M_H^2,M_Z^2,M_Z^2\right) }{64 (D-2) \pi ^2 \left(t-4
   M_H^2\right){}^2 M_W^2 \left(M_W^2-M_Z^2\right) \left(4
   M_Z^2-M_H^2\right) M_H^2} 
   \biggl[
   \left((20-8 D)
   M_H^2+3 (D-2) t\right) M_H^8 \nonumber\\
   &+& 4 \left(16 \left(D^2-4 D+3\right) M_H^4-4
   \left(D^2-3 D+2\right) t M_H^2-\left(D^2-3 D+2\right) t^2\right) M_Z^6 - 4
   \left(4 \left(2 D^2-3 D-7\right) M_H^6 \right. \nonumber\\  
   &+& \left. \left(18-3 D^2+5 D\right) t
   M_H^4  + 3 (D-2) t^2 M_H^2\right) M_Z^4 +\left(16 (3 D-8) M_H^8-16 (D-3) t
   M_H^6+3 (D-2) t^2 M_H^4\right) M_Z^2
   \biggr] \nonumber\\
   &+& \frac{e^2 M_Z^2
   \left(M_H^4-4 M_Z^2 M_H^2+4 (D-1) M_Z^4\right) A_0\left(M_Z^2\right)}{64
   \pi ^2 M_W^2 \left(M_W^2-M_Z^2\right) \left(4 M_Z^2-M_H^2\right)
   M_H^2} 
   \,.
\label{theta2Explicit}
\end{eqnarray}

%
%
%
%
%
%
%

\end{document}